\documentclass{article}

\usepackage{PRIMEarxiv}
\usepackage[numbers]{natbib}
\RequirePackage{amsthm,amsmath,amsfonts,amssymb}

\usepackage[utf8]{inputenc} % allow utf-8 input
\usepackage[T1]{fontenc}    % use 8-bit T1 fonts

\usepackage{hyperref}       % hyperlinks
\usepackage{url}            % simple URL typesetting
\usepackage{booktabs}       % professional-quality tables
\usepackage{amsfonts}       % blackboard math symbols
\usepackage{nicefrac}       % compact symbols for 1/2, etc.
\usepackage{microtype}      % microtypography
\usepackage{lipsum}
\usepackage{fancyhdr}       % header
\usepackage{graphicx}       % graphics
\graphicspath{{media/}}     % organize your images and other figures under media/ folder
\newcommand{\ve}[1]{{\mbox{\boldmath ${#1}$}}}  
\usepackage[onehalfspacing]{setspace}

\usepackage{stmaryrd}
\title{A canonical polyadic tensor basis for fast Bayesian estimation of multi-subject fMRI activation patterns
}

\author{
  Michelle F. Miranda \\
  Department of Mathematics and Statistics \\
  University of Victoria \\
  Victoria, BC, Canada\\
  \texttt{michellemiranda@uvic.ca} \\
}

\begin{document}
\maketitle

\begin{abstract}
Task-evoked functional magnetic resonance imaging studies, such as the Human Connectome Project (HCP), are a powerful tool for exploring how brain activity is influenced by cognitive tasks like memory retention, decision-making, and language processing. A fast Bayesian function-on-scalar model is proposed for estimating population-level activation maps linked to the working memory task. The model is based on the canonical polyadic (CP) tensor decomposition of coefficient maps obtained for each subject. This decomposition effectively yields a tensor basis capable of extracting both common features and subject-specific features from the coefficient maps. These subject-specific features, in turn, are modeled as a function of covariates of interest using a Bayesian model that accounts for the correlation of the CP-extracted features. The dimensionality reduction achieved with the tensor basis allows for a fast MCMC estimation of population-level activation maps. This model is applied to one hundred unrelated subjects from the HCP dataset, yielding significant insights into brain signatures associated with working memory.

\end{abstract}

% keywords can be removed

\section{Introduction}
Task-evoked functional magnetic resonance imaging (fMRI) studies, such as the Human Connectome Project (HCP), are a powerful tool for exploring how brain activity is influenced by cognitive tasks such as memory retention, decision-making, and language processing. In the HCP, a core set of tasks was devised aiming to delineate the functionally unique nodes within the human brain to establish connections between tasks and activation signatures in critical network nodes \citep{Barch2013}. These data have motivated the work presented here.

Brain imaging data is often represented in the form of a multi-dimensional array, called a tensor. These arrays can be seen as functions of space and/or time. In multi-subject studies, it is of interest to relate these functional response data to a set of scalar predictors through a function-on-scalar model. Due to the high-dimensionality of the functional response, most models in the literature will employ a form of regularization. Regularization involves borrowing strength across observations within a function. Although regularization can be achieved through various methods such as penalized regression methods or/and sparse-inducing priors, one very important approach is by the use of basis functions. Each basis function defines a linear combination among the locations within the function, establishing a framework for borrowing strength between observations. Commonly used basis functions are splines, Fourier series, wavelets, and principal components (PCs)\citep{Morris2015}.

Some efforts have been devoted to analyzing neuroimaging as tensor data. In one line of research, the CP decomposition is used as a descriptive tool to represent multisubject and multisession fMRI data. The resulting spatial patterns or modes of the tensor representation are, in turn, validated by visual inspection. The extracted components in each spatial, temporal, runs, among other dimensions, can be further inspected to describe variability in each of these directions \citep{ANDERSEN2004728}. Although this descriptive decomposition is useful, it is often of interest to relate the tensor neuroimaging data with other variables such as age, gender, and sex, disease status, through a regression analysis framework.

Previous research that tackles tensor regression are models in which the response is a scalar and the predictor is a tensor \citep{Zhu2013,Miranda2018,Li2018,Guhaniyogi2017}. Commonly in these approaches, the goal is to understand the changes of the response outcome as the tensor variable changes. Often,
a tensor decomposition is implemented as tool for dimensionality reduction of tensor covariates. For example, the authors in \cite{Miranda2018} use this approach to predict Alzheimer's Disease status as a function of magnetic resonance imaging data through a CP decomposition of the imaging covariate.

Meanwhile, tensor response regression aims to study the change of the image as the predictors such as the disease status and age vary. Research in this direction often propose a low-rank tensor decomposition of the coefficients associated with covariates \citep{Spencer2020, rabusseau2016low}. In a different application context, \cite{Lock2018} propose a likelihood-based latent variable tensor response regression, where the tensor data corresponds to latent variables informed by additional covariates. The authors apply their model to a 2D facial image database with facial descriptor.

Although aiming to address the same type of problem, our proposal is methodologically distinct from these previous approaches. Here, we use the CP tensor decomposition to construct a tensor basis function that not only determines the mechanisms for borrowing information across voxels but also effectively reduces the dimensionality of the brain data, which is projected onto the basis space. The projected data is in turn modeled as functions of covariates of interest using a Bayesian MCMC algorithm. This approach is inspired by the wavelet functional mixed models proposed in \cite{Morris06} and further developed in \citep{Morris11,Zhang2016,Zhu2011,Zhu2018}. The novelty of the methodology here is that the CP decomposition captures the 3D spatial brain intricacies across different subjects and concatenation of the spatial directions does not happen as in wavelet basis or principal components. We apply this approach to a second-level analyses of fMRI data from the one hundred unrelated subjects of the HCP data. The method, however, extends to a vast range of imaging modalities such as structural MRI images, and electroencephalography data.

The rest of the paper is organized as follows. In Section \ref{sec:methods} we introduce the CP decomposition and estimation method, including a brief review of tensor concepts and properties (Section \ref{sec:background}) and a review of the method used to obtain subject-level brain maps for fMRI task data (Section \ref{subsec:method:subjectlevel}, and the design the CP tensor basis functions with the Bayesian modeling approach (Section \ref{sec:methodmulti}). In Section \ref{sec:app} we apply the proposed approach to the Working Memory data from HCP to find population-level maps of a particular Working Memory contrast. In Section \ref{sec:discc} we provide some concluding remarks and directions for future work.

%\newpage
\section{Methods}
\label{sec:methods}
The proposed approach involves a two-stage procedure. In the first stage, the 4D data for each subject are modeled through the Bayesian composite-hybrid basis model proposed in \citep{Miranda2021} and described in Section \ref{subsec:method:subjectlevel}. In the second stage, posterior means of contrast brain maps are modeled by the CP tensor basis model. This modeling strategy is flexible, as subject-level maps can be estimated by any known single-subject estimation technique and combined into coefficients maps for any contrasts of interest. Furthermore, these maps are assembled into a 4D tensor (3D volume and subjects) to be associated with population-level covariates, e.g., age and sex.

\subsection{Background}
\label{sec:background}

We review a few important concepts and operations necessary for the comprehension of this manuscript. For an exhaustive review on tensor decomposition and applications, see \cite{kolda2009tensor}.

\subsubsection{Working with tensors}

A {\itshape tensor} is a multidimensional array with order given by the number of its dimensions, also known as modes. For example, a matrix is a tensor of order 2. A tensor $\mathcal{X} \in \mathbb{R}^{I_1 \times I_2 \ldots \times I_K}$ of order $K$ is said to be a {\itshape rank-one tensor} if it can be written as an outer product of $K$ vectors, i.e.,

\begin{equation}
\mathcal{X}=\ve a^{(1)} \circ \ve a^{(2)} \circ \ldots \circ \ve a^{(K)},
\end{equation}

where $\circ$ is the outer product of vectors. This is equivalent to writing each element of $x_{i_1i_2\ldots i_K}= a^{(1)}_{i_1} a^{(2)}_{i_2} \ldots a^{(K)}_{i_K}$, for $1 \leq i_k \leq i_K$.

{\itshape Matricization or unfolding} of a tensor is the process of reordering the tensor into a matrix. We will denote the mode-$n$ matricization of the tensor $\mathcal{X}$ above as $\ve X_{(n)}$ and arrange the mode-$n$ dimension to be the columns of $\ve X_{(n)}$. Please see \cite{kolda2009tensor} for how the observations are arranged for a general tensor order. As an example, consider the matricization of a tensor $\mathcal{X} \in \mathbb{R}^{2 \times 3 \times 4}$. The mode-$3$ matricization $\ve X_{(3)}$ is a matrix of size $6 \times 4$ where, for a fixed column $k$ in the matrix, we have elements $x_{11k},x_{21k},x_{x12k},x_{22k},x_{13k},x_{23k}$ in the $k$th row.

\begin{figure}[h!]
\begin{center}
\includegraphics[width=0.8\textwidth]{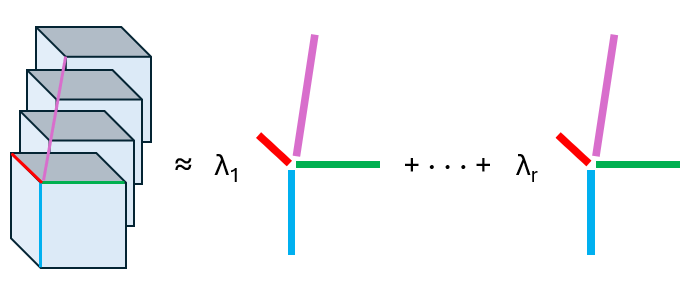} % This is a *.eps file
\end{center}
\caption{CP decomposition of a 4D array into the sum of rank-one tensors.}
\label{fig:decomp}
\end{figure}

{\itshape Tensor decompositions} are higher-order generalizations of the singular value decomposition and principal component analysis. Although there are various tensor decomposition methods, we focus here on the Canonical Polyadic (CP). The CP decomposition factorizes a tensor as a sum of rank-one tensors, as illustrated in Figure \ref{fig:decomp} for a tensor of order four.

Mathematically, we write the CP decomposition above as

\begin{equation}
\mathcal{X} \approx \llbracket \ve \lambda; \ve A,\ve B, \ve C,\ve D \rrbracket \equiv \sum_{r=1}^R \lambda_r \ve a_r \circ \ve b_r \circ \ve c_r \circ \ve d_r
\end{equation}

The factorization means that every element of the four-dimensional array $\mathcal{Y} \in \mathbb{R}^{J_1 \times J_2 \times J_3 \times J_4} $ can be written as $y_{j_1j_2j_3j_4}=\sum_{r=1}^R \lambda_r a_{j_1r} b_{j_2r}c_{j_3r}d_{j_4r}$, where the vectors $\ve a_r \in \mathbb{R}^{J_1}$, $\ve b_r \in \mathbb{R}^{J_2}$, $\ve c_r \in \mathbb{R}^{J_3}$, and $\ve d_r \in \mathbb{R}^{J_4}$ form the columns of the factor matrices $\ve A, \ve B, \ve C,$ and $\ve D$.

The {\itshape factor matrices} refer to the combination of the vectors from the rank-one components, i.e., $\ve A = [\ve a_1 \ve a_2 \ldots \ve a_R]$ and likewise for B, C, and D. We assume that the columns of A, B, C, and D are normalized to length one, with the weights absorbed into the vector $\ve \lambda \in \mathbb{R}^R$. The approximation refers to the estimation of the factor matrices that minimize the Frobenius norm of the difference between $\mathcal{X}$ and $\llbracket \ve \lambda; \ve A,\ve B, \ve C,\ve D \rrbracket$ \citep{carroll1970analysis}.

\subsection{Subject-level Modeling}
\label{subsec:method:subjectlevel}

For each voxel $v=1,\ldots,N_v$ in a volumetric image, the task fMRI response function is a time series of $T$ time points based on the Blood Oxygen Level Dependent (BOLD) contrast \citep{Ogawa1990}. These measurements are an indirect and non-invasive measure of brain activity and can be modeled as a function of the external stimuli:

\begin{equation}
\label{eq:voxelwise}
  y_t(v) = \sum_{j=1}^p b_j(v)x_j(t) + e_t(v),  
\end{equation}

where $x_j(t)$ is the covariate for stimulus $j$, representing the indicator function of stimulus $j$ at time $t$ convolved with the hemodynamic response of a neural event at time $t$ \citep{lindquist2008}, and $e_t(v)$ is a measurement error. The coefficients $b_j(v)$ characterize the relationship of brain activity during stimulus $j$. The primary goal of activation studies is to identify which voxels are differentially activated by specified stimuli.

There are many ways to estimate the coefficient maps $b_j(v)$, including the popular generalized linear model available through the software SPM \citep{friston2006statistical}. The model in \eqref{eq:voxelwise} can be written in matrix form as:

\begin{equation}
      \mathbb{Y} = \mathbb{X} \ve B + \mathbb E
\end{equation}

To estimate the brain maps $\ve B \in \mathbb{R}^{p \times N_v}$, we will use the Bayesian approach proposed in \cite{Miranda2021}, which consists of fitting a basis-space model:

\begin{equation}
      \mathbb{Y^*} = \mathbb{X^*} \ve B^* + \mathbb E^*
\end{equation}

This can be written in vectorized form as:

\begin{equation}
 \text{vec}(\mathbb Y^*) = (\ve I_S \otimes \mathbb X^*)( \ve I_S \otimes \ve B^*) + \text{vec}(\mathbb E^*),
\end{equation}

with $\text{vec}(\mathbb Y^*) = (\ve\Psi^T \ve\Phi^T \otimes \ve W) \text{vec}({\mathbb Y})$ and $\text{vec}(\mathbb E^*) = (\ve\Psi^T \ve\Phi^T \otimes \ve W) \text{vec}(\mathbb E)$. In summary, the authors' tensor basis approach consists of choosing the set of bases $\{\ve \Psi, \ve \Phi, \ve W\}$ to account for global and local spatial correlations and long-memory temporal correlations. A Bayesian MCMC is proposed and samples from the coefficient maps are obtained in the basis space, and the inverse transform is applied to obtain the MCMC samples in the voxel space. For more details, refer to \cite{Miranda2021}.

\subsection{Multi-subject Modeling}
\label{sec:methodmulti}

Let $\mathcal{Y}$ be the 4D tensor representing each coefficient map or contrast map for all subjects. Then $\mathcal{Y} \in \mathbb{R}^{p_1 \times p_2 \times p_3 \times N}$. We can model $\mathcal{Y}$ as a function of $p$ subject-level covariates $\ve Z \in \mathbb{R}^p$ as follows:

\begin{equation}
\label{eq:modelbetaspace}
\ve Y = \ve Z \ve \gamma + \ve E,
\end{equation}

where $\ve Y$ is a $N \times N_v$ matrix representing the mode four unfolding of the tensor $\mathcal{Y}$; $\ve Z$ is a $N \times p$ design matrix of subject-level covariates; $\ve \gamma$ is a $p \times N_v$ matrix of coefficients; and $\ve E$ is a $N \times N_v$ matrix of random errors.

Properly modeling the data as in \eqref{eq:modelbetaspace} is challenging due to its very large number of voxels, typically around a million, and the spatial dependency inherited from the brain volume data. Most Bayesian models rely on either downsampling the volume data to reduce the number of voxels or rely on simplified assumptions on the prior of $\ve \gamma$ and/or $\ve E$. The solution proposed here involves projecting the data into a basis space through a tensor basis obtained from the CP decomposition of the tensor $\mathcal{Y}$. This decomposition is low rank and effectively extracts brain volume features that are common to all subjects. By extracting the common features, the remaining information is subject-specific and can be then related to the covariates of interest.

The CP decomposition of $\mathcal{Y}$ is given by:

\begin{equation}
\label{eq:cpdata}
   \mathcal{Y} \approx \sum_{r=1}^R \lambda_r \ve a^{(1)}_r \circ \ve a^{(2)}_r \circ \ve a^{(3)}_r \circ \ve g_r \equiv \llbracket \ve \lambda; \ve A^{(1)},\ve A^{(2)},\ve A^{(3)},\ve G \rrbracket,
\end{equation}

where the vectors $\ve a^{(1)}_r \in \mathbb{R}^{p_1}$, $\ve a^{(2)}_r \in \mathbb{R}^{p_2}$, $\ve a^{(3)}_r \in \mathbb{R}^{p_3}$, and $\ve g_r \in \mathbb{R}^{N}$. The matrices $\ve A^{(1)}$, $\ve A^{(2)}$, $\ve A^{(3)}$, and $\ve G$ are called factor matrices and their columns are associated with these vectors. From \cite{kolda2009tensor}, we can write $\ve Y$ from model \eqref{eq:modelbetaspace} as:

\begin{equation}
\ve Y = \ve G \ve \Lambda (\ve A^{(3)} \odot \ve A^{(2)}  \odot \ve A^{(1)})^T + \ve \epsilon,
\end{equation}

where $\ve G$, $\ve A^{(1)}$, $\ve A^{(2)}$, and $\ve A^{(3)}$ are estimated to minimize the Frobenius norm of $\ve \epsilon$ \citep{kolda2009tensor}. Let $\ve L = \ve \Lambda (\ve A^{(3)} \odot \ve A^{(2)}  \odot \ve A^{(1)})^T \in \mathbb{R}^{N_v \times R}$, then $\ve P = \ve L^{\dagger}$, the Moore-Penrose pseudoinverse of $\ve L$, provides a projection matrix formed by spatial basis that can be used to effectively reduce data from $N_v$ to $R$ spatial locations.

Based on the projection matrix, we multiply both sides of equation \eqref{eq:modelbetaspace} by $\ve P$ to obtain the basis space model:

\begin{equation}
\label{eqbasisspace}
    \ve G = \ve Z \ve \gamma^* + \ve E^* ,
\end{equation}

where $\ve \gamma^* = \ve \gamma \ve P$ is the $p \times R$ matrix of coefficients in the basis space, and $\ve E^* = \ve E \ve P$ is a $N \times R$ matrix of random errors.

\subsubsection{Basis-space Bayesian Modeling}
\label{sub:mcmc}

The columns of $\ve G$ are correlated due to the nature of the CP decomposition, and model assumptions should reflect that. For each row of $\ve E^*$, we assume $\ve E_i^* \sim N(0,\ve \Sigma_{\epsilon})$ and priors:

\begin{equation}
    \ve p(\ve \gamma^*, \ve \Sigma_{\epsilon}) = \ve p(\ve \Sigma_{\epsilon})\ve p(\ve \gamma^*/\ve \Sigma_{\epsilon}),
\end{equation}

with $\ve \gamma^* \sim \text{MN}(\ve g_0, \ve L_0^{-1},\ve\Sigma_{\epsilon})$ and $\ve \Sigma_{\epsilon} \sim \text{W}^{-1}(\ve V_0,\nu_0)$, where $\text{MN}$ indicates the matrix normal distribution and $\text{W}^{-1}$ indicates the Inverse Wishart distribution. The joint posterior distribution of $\ve \gamma^*, \ve \Sigma_{\epsilon}/ \ve G, \ve Z$ factorizes as a product of an Inverse Wishart and a Matrix Normal distributions as follows:

\begin{equation*}
 \ve \Sigma_{\epsilon}/ \ve G, \ve Z   \sim \text{W}^{-1}(\ve V_n,\nu_n),
\end{equation*}
\begin{equation}
 \ve \gamma^*/ \ve G, \ve Z,\ve \Sigma_{\epsilon}   \sim \text{MN}(\ve g_n,\ve L_n^{-1},\ve \Sigma_{\epsilon}),
\end{equation}

with:

\[\ve V_n = \ve V_0 + (\ve Y - \ve Z \ve g_n)^T (\ve Y - \ve Z \ve g_n) + (\ve g_n - \ve g_0)^T\ve L_0(\ve g_n - \ve g_0),\]
\[\nu_n = \nu_0 + N,\]
\[\ve g_n = (\ve Z^T\ve Z + \ve L_0)^{-1}(\ve Z^T \ve G + \ve L_0\ve g_0),\]
\[\ve L_n = \ve Z^T \ve Z + \ve L_0.\]

Sampling from these distributions is straightforward and done through Gibbs sampling.

\subsubsection{Rank Selection}
\label{subsec:rank}

To choose the appropriate rank for the CP decomposition, we propose a 10-fold cross-validation procedure with the following steps for each of the training/test set combination:

\begin{enumerate}
    \item Obtain the factor matrices of the CP decomposition for the training data,
    \item Proceed to estimate $\ve \gamma^*$ from model \eqref{eqbasisspace} following the MCMC algorithm detailed in Section \ref{sub:mcmc},
    \item Using $\widehat{\ve {\gamma}^*}$ from the previous step and $\ve Z_{\text{Test}}$, estimate $\ve G_{\text{Test}} = \ve Z_{\text{Test}} \widehat{\ve {\gamma}^*}$,
    \item Obtain $\widehat{\mathcal{Y}}_{\text{Test}} = \llbracket \ve \lambda; \ve A^{(1)},\ve A^{(2)},\ve A^{(3)},\ve G_{\text{Test}} \rrbracket$ as in \eqref{eq:cpdata},
    \item Choose the rank that minimizes the Frobenius norm of the average of the $\mathcal{Y}_{\text{Test}} - \widehat{\mathcal{Y}}_{\text{Test}}$, obtained across the training/test sets.
\end{enumerate}

\subsubsection{Inference on the Brain Space}

For every MCMC sample of $\ve \gamma^*$, we  obtain $\ve \gamma= \ve \gamma^*\ve L$ in the original voxel space. To control for the experimentwise error rate in the voxel space, joint credible bands as in \cite{Ruppert2003} can be calculated. For each voxel $v$, a joint $100(1-\alpha)\%$ credible interval for any group-level contrast $\ve C$ is given by
\begin{equation}
    I_{\alpha}(v)=\hat{C}(v) \pm q_{(1-\alpha)}[\widehat{\mbox{std}}(\hat{C}(v))],
\end{equation}
 where the quantile $q_{(1-\alpha)}$ is obtained from $z^{(1)},\ldots, z^{(M)}$ with $z^{(m)}=\max_{v\in \mathcal{V}}\left|\frac{C^{(m)}(v)-\hat{C}(v) }{\widehat{\mbox{std}}(\hat{C}(v))} \right|$, for $m=1,\ldots, M$ and $M$ is the total number of MCMC samples after burn-in and thinning \cite{Miranda2021}.

\noindent These joint credible bands account for multiple testing across the voxel space and can be inverted to flag activation signatures based on voxels for which the $100(1-\alpha)\%$ joint credible bands exclude zero \citep{Meyer2015, Miranda2021}. Denote by $P_{SimBas}$ the minimum $\alpha$ at which each interval excludes zero ($P_{SimBas}=\min\{\alpha: 0 \notin I_{\alpha}(v)\}$). $P_{SimBas}$ can be directly computed by
\begin{equation}
\label{eq:simbas}   
P_{SimBas}(v)=\frac{1}{M}\sum_{m=1}^M1\left\{\left|\frac{\hat{C}(v)}{\widehat{\mbox{std}}(\hat{C}(v))}\right|\leq z^{(m)}\right\}.
\end{equation}

\section{Application to the HCP data}
\label{sec:app}

The proposed model is applied to the 100 unrelated subjects data from the Working Memory task of the Human Connectome Project. The volumes considered were collected from the right-left phase. The experiment consists of a total of 8 blocks, half corresponding to the 2-back task and the other half the 0-back task. In each block, the  participants  were shown a series of 10 images. Images corresponding to 4 stimuli (tools, places, body parts, faces) were embedded within the memory task. Each image was shown for 2.5s, followed
by a 15s inter-block interval. fMRI volumes were obtained every 720 ms (TR). Each volume consisted of images of size $91 \times 109 \times 91$ for a total of 405 time frames. 

{\bf Subject-level analysis}. We follow the analysis described in Section \ref{subsec:method:subjectlevel}. To obtain the local spatial basis, we partitioned the brain into ROIs using a digitalized version of the original Talairach structural labelling that was registered into the  MNI152 space. The atlas can be obtained from the FSL atlas library with information found at {\url{ https://fsl.fmrib.ox.ac.uk/fsl/fslwiki/Atlases}}  \citep{Jenkinson2012}. We considered ROIs that had at least 125 voxels and obtained a total of 298 regions. Within  each region, the local basis $\ve \Phi$ was formed by the eigenvectors of the SVD decomposition for the matrix data in each ROI as described in \cite{Miranda2021}. Next, the global spatial basis $\ve \Psi$ was obtained by first projecting the brain data using the local basis and then calculating the SVD decomposition of the project data, exactly as in \cite{Miranda2021}. The temporal basis $\ve W$ was chosen to be the Haar wavelet. The estimation was performed at the cluster Cedar managed by the Digital Research Alliance of Canada (former Compute Canada). The jobs were submitted in parallel for the 100 subjects and results took an hour to finish.

For each subject, after estimation, we constructed a contrast map of the 2 back versus 0 back working memory task. The maps were obtained from the posterior means of the coefficients, and assembled into a 4D array $\mathcal{Y} \in \mathbb{R}^{91 \times 109 \times 91 \times 405}$.
%^{91 \times 109 \times 91 \times 405}$

\begin{figure}[b]
\begin{center}
\includegraphics[width=14cm]{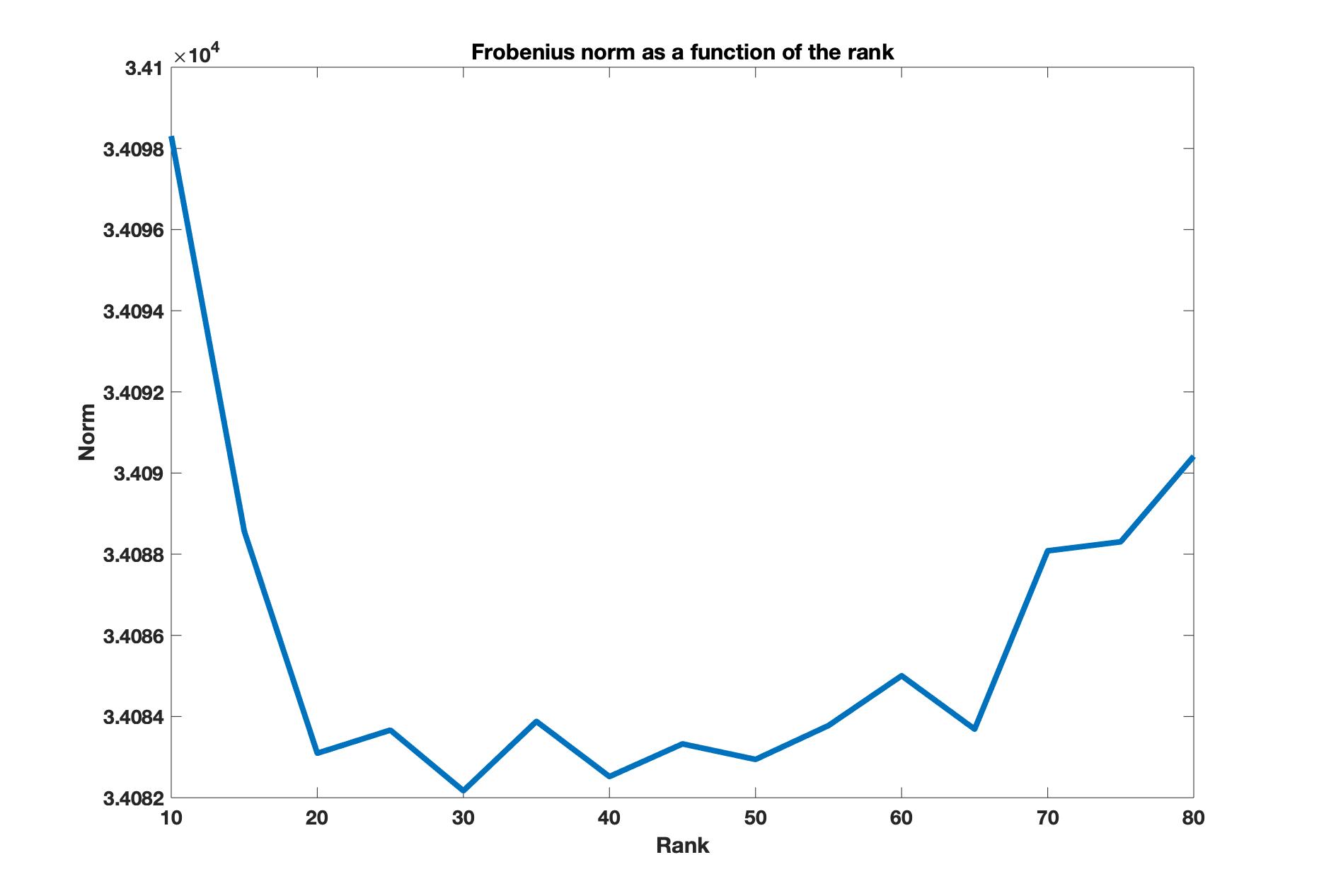}% This is a *.eps file
\end{center}
\caption{Cross-validation to choose the rank R. We want R to minimize the Frobenius norm.}\label{fig:rank}
\end{figure}

{\bf Multi-subject analysis}. We take the tensor $\mathcal{Y}$ containing the posterior means of the contrast 2back versus 0back for the maps obtained in the single-level analysis. Next, we follow the method proposed in Section \ref{sec:methodmulti} to estimate the coefficients maps $\ve \gamma$ for the mean of 2back versus 0back and the covariate {\itshape Sex}. We are interested in finding common activation maps for all subjects and aim to investigate if there are differences in working memory task based on sex. Next, we  obtain the CP decomposition of the tensor $\mathcal{Y}$, taking $R=30$ selected based on the cross-validation procedure described in Section \ref{subsec:rank}. Figure \ref{fig:rank} shows the Frobenius norm as a function of the rank, for rank values $R=10, 15, 20, 25, \ldots, 80$. Following the decomposition, we aim to identify significant locations in the brain for the Working Memory task 2back versus 0 back, we proceed as follows.

\begin{enumerate}
    \item[(i)] Obtain MCMC samples for $\ve \gamma^*$, as detailed in Section \ref{sub:mcmc}.
    \item[(ii)] Use the inverse transform of $\ve P$, $\ve L=\ve \Lambda (\ve A^{(1)} \odot \ve A^{(2)}  \odot \ve A^{(3)})^T \in \mathbb{R}^{N_v \times R}$, to obtain MCMC samples for $\ve \gamma=\ve \gamma^* \ve L$ in the voxel space.
    \item[(iii)] For each subject-level covariate, use the MCMC samples of $\ve \gamma$ to obtain $P_{SimBas}$ as described in \eqref{eq:simbas} to control for the experimentwise error rate on multiple testing.
    \item[(iv)] For $P_{SimBas}<\alpha=0.01$ flag the voxel as different than zero, indicating activation.
    \item[(v)] Obtain clusters of significant voxels that are at least 125 voxels large.
\end{enumerate}

 The computational time to run this analysis is 1.62 minutes on a MacBook Pro 2.8 GHz Quad-Core Intel Core i7 with 16 GB 2133 MHz LPDDR3 memory. For the cross-validation procedure, it took 38 minutes for a grid of 15 rank values. For faster computational time on rank selection, one could run a five-fold cross validation instead and/or a less fine grid of ranks.

\begin{figure}[h!]
\begin{center}
\includegraphics[width=13cm]{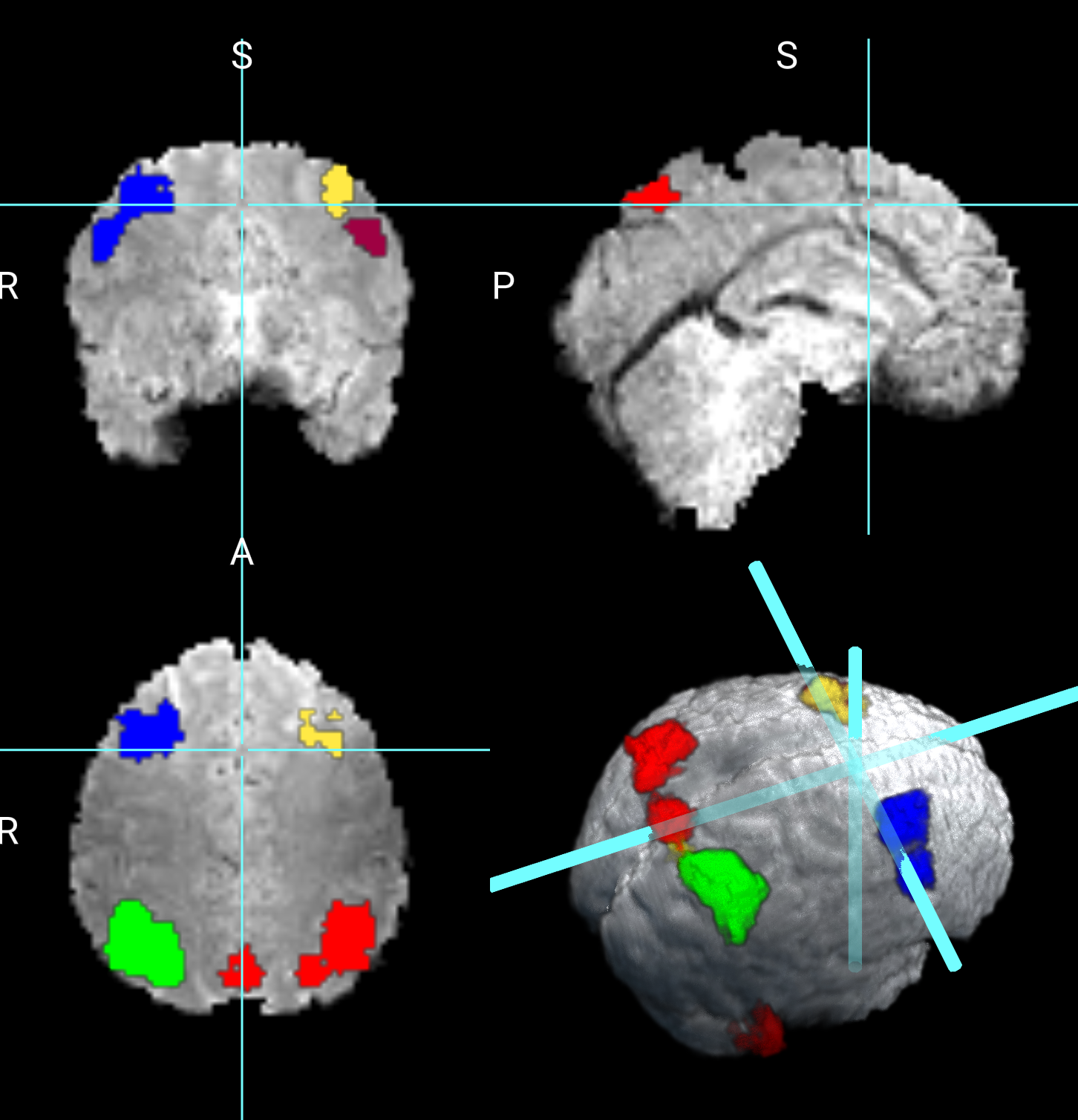}% This is a *.eps file
\end{center}
\caption{Working Memory contrast of 2back versus 0back.}\label{fig:average}
\end{figure}

We consider two maps of coefficients, the first is the intercept, which is the average activation for the contrast 2 back versus 0 back of the Working Memory task. The second is the map associated with the variable sex. We did not find any significant differences in the working memory contrast between females and males.   Figure \ref{fig:average} shows the results for the intercept revealing 4445 active voxels. The larger cluster in red shows 1268 voxels encompassing 24 distinct brain regions as subdivided by the Tailarach Atlas,  (labels can be found here \url{http://www.talairach.org/about.html#Labels}). Within this cluster, larger active regions are in the right parietal lobe, including inferior, superior, precuneous parietal regions. The cluster in green is formed by 1182 voxels, with active regions on the left parietal lobe, including the inferior, superior, precuneous, and cuneous regions. The parietal lobe has been previously identified as playing a key role in working memory \citep{BERRYHILL20081767, LEE2022126,Koenigs14980}. 

The next cluster in blue shows the right frontal lobe, including the superior and middle frontal gyrus. The cluster in yellow shows 345 voxels in the right frontal lobe, including the superior and middle frontal gyrus. The cluster in light orange shows 276 voxels on the left frontal lobe in the middle frontal gyrus. These regions have also been identified as playing a critical role in working memory \citep{Owen1998}. The cluster in burgundy indicates 313 voxels on the right cerebellum at the posterior lobe region.  A smaller cluster of 139 voxels was also identified on the posterior lobe at the left cerebellum. The authors in \citep{deverett2019cerebellar} found that the cerebellum can influence the accurate maintenance of working memory.

\begin{figure}[h!]
\begin{center}
\includegraphics[width=13cm]{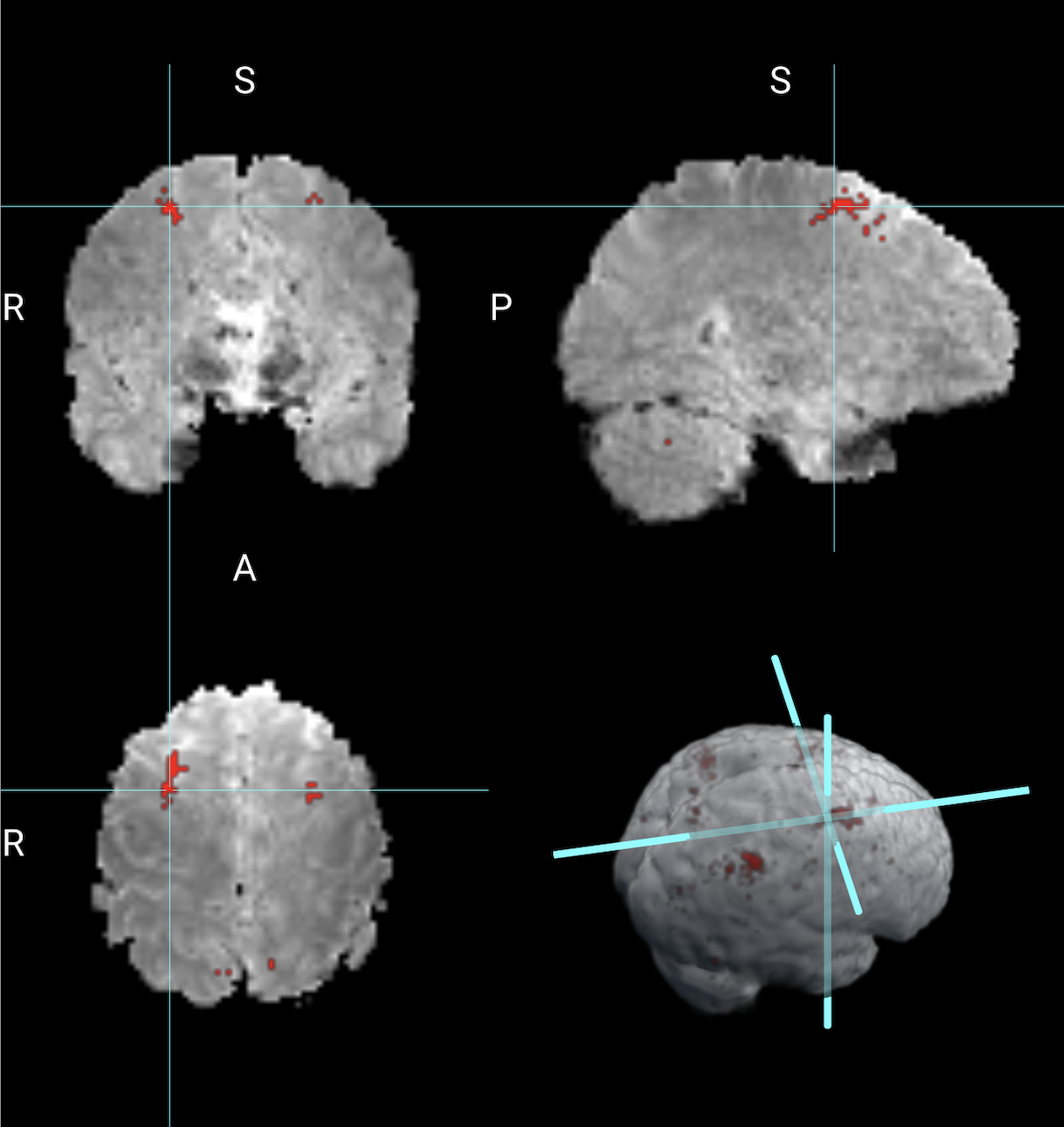}% This is a *.eps file
\end{center}
\caption{Working Memory contrast of 2back versus 0back obtained from SPM.}\label{fig:average_spm}
\end{figure}

Next, we compare the results obtained from the Bayesian CP tensor basis approach with the results obtained from the Statistical Parametric Map (SPM) software \citep{friston2006statistical}. We ran both the first and the second-level analysis in SPM with parameter settings to reflect no spatial normalization and an AR(1) noise component. The results for the Average of 2 back versus 0 back contrast are displayed in Figure \ref{fig:average_spm}. The two largest clusters were sizes 64 and 58 voxels, respectively. Overlaid onto the template is a mask with all significant voxels based on a voxel-level threshold that controls for the family-wise error rate of 0.05. Inspecting Figure \ref{fig:average_spm} reveals that this SPM voxelwise model only capture small parts of the parietal and frontal lobe and has significant less power than the CP tensor basis approach.

\section{Final Remarks}
\label{sec:discc}
In this paper, we propose a fast Bayesian model that takes tensor responses as a function of population-level covariates. The model is based on using the CP tensor decomposition approach to construct a basis function with the goal of modeling the spatial structure of the tensor data. This approach allows us to borrow information across voxels, increasing the power for the statistical tests performed. Because the estimation of the modeling components is performed in the basis space, we have a very fast Bayesian procedure. The Bayesian MCMC estimation method for the second-level analysis, including cross-validation for rank selection, estimation, and inference, takes 40 minutes on a regular laptop. This time can be much faster if using a cluster with more powerful computing capabilities.

We apply this approach to a second-level analysis of fMRI data from one hundred unrelated subjects of the HCP dataset. The results obtained are consistent with findings in the literature showing the role of the parietal lobe, the middle frontal gyrus, and the posterior lobe of the cerebellum in working memory.

One limitation of the proposed approach is that CP decomposition applies only to volumetric data. For more modern imaging files, such as the CIFTI files available in the Human Connectome Project, the CP decomposition would need to aggregate spatial information as one large dimension. In this case, the benefits of the CP decomposition are unclear as the CP decomposition might be similar to other basis approaches, including principal components and wavelets. The authors plan to conduct an investigation on that front.

\section*{Funding}
This research was funded by Natural Sciences and Engineering Research Council of Canada grant number RGPIN-2020-06941.

%\section*{Acknowledgments}
%This is a short text to acknowledge the contributions of specific colleagues, institutions, or agencies that aided the efforts of the authors.

%\section*{Supplemental Data}
% A MATLAB toolbox 

\bibliographystyle{plain} %  Many Frontiers journals use the Harvard referencing system (Author-date), to find the style and resources for the journal you are submitting to: https://zendesk.frontiersin.org/hc/en-us/articles/360017860337-Frontiers-Reference-Styles-by-Journal. For Humanities and Social Sciences articles please include page numbers in the in-text citations 
\bibliography{test}

%%% Make sure to upload the bib file along with the tex file and PDF
%%% Please see the test.bib file for some examples of references

\end{document}